# REVISIÓN LONGITUDINAL DE PORTAFOLIOS CON ENFOQUE DE MÍNIMA VARIANZA ANTES, DURANTE Y DESPUÉS DE LA PANDEMIA

*Longitudinal review of portfolios with minimum variance approach before during and after the pandemic*

**Genjis A. Ossa González**
Universidad Popular del Cesar (Colombia)
gossa@unicesar.edu.co

**Luis H. Restrepo Sierra**
Universidad Popular del Cesar (Colombia)
luisrestrepo@unicesar.edu.co

**Resumen**

Este estudio investiga el impacto de la pandemia en las acciones con mayor transacción en el mercado de valores de Colombia para la fecha del 17 de enero de 2024. Basándonos en los datos diarios de las empresas más transaccionadas en Colombia para dicha fecha y abarcando un período general de 2015 a 2023, de manera sintetizada nuestro análisis revela que en el periodo 2015-2019, el rendimiento alcanzó un 5,70%, con un riesgo relativamente bajo del 18,45%. Sin embargo, en el siguiente periodo 2016 -2020, a pesar de que el rendimiento disminuyo a 5,40%, el riesgo experimentó un aumento significativo, llegando al 24,64%. La beta también mostró variaciones, siendo más baja en 2015-2019 con 0,61 y aumentando a 1,02 en 2016-2020. La línea de mercado de capitales (LMC) en los portafolios construidos tienen tendencia descendente, indicando que la cartera ofrece una tasa de rendimiento esperada inferior a la tasa libre de riesgo. Este hallazgo se respalda en el índice de Sharpe, que muestra valores negativos a lo largo de los períodos estudiados.

Palabras Clave: Frontera eficiente; CAPM; Colombia; Portafolio, Rendimientos.

# Abstract

This study investigates the impact of the pandemic on the most traded stocks in the Colombian stock market for the date of January 17, 2024. Based on the daily data of the most traded companies in Colombia for said date and covering a period general from 2015 to 2023, in a summarized way our analysis reveals that in the period 2015-2019, the return reached 5.70%, with a relatively low risk of 18.45%. However, in the following period 2016 -2020, although the yield decreased to 5.40%, the risk experienced a significant increase, reaching 24.64%. The beta also showed variations, being lowest in 2015-2019 with 0.61 and increasing to 1.02 in 2016-2020. The capital market line (LMC) in the constructed portfolios has a downward trend, indicating that the portfolio offers an expected rate of return lower than the risk-free rate. This finding is supported by the Sharpe index, which shows negative values throughout the periods studied.

## 1. INTRODUCCIÓN

La pandemia de COVID-19 ha dejado una marcada huella en los mercados financieros, generando niveles sin precedentes de incertidumbre. En este contexto, el dinámico escenario financiero actual y la toma de decisiones de inversión requiere una comprensión profunda de las tendencias del mercado y la estrategia de construcción de portafolios bajo el enfoque de mínima varianza se vuelve esencial para la gestión del riesgo y la protección del capital.

La Teoría de Carteras Moderna (TCM) desarrollado por Markowitz, H. (1952) que considera el equilibrio entre rendimiento y riesgo sostiene que los inversores pueden construir carteras óptimas que maximizan el rendimiento esperado para un nivel de riesgo dado o minimizan el riesgo para un rendimiento (Kolm et al., 2014), este último [mínima varianza] busca la selección de activos que minimicen la volatilidad global del portafolio. Esta estrategia se fundamenta en la premisa de que los inversores racionales buscan maximizar la rentabilidad considerando el riesgo, siendo la volatilidad una medida crucial de este último (Chaweewanchon, A., y Chaysiri, R. 2022).

Markowitz, H. (1952) introdujo la matriz de covarianza como herramienta fundamental para medir la relación entre los rendimientos de diferentes activos, siendo esencial en los cálculos de varianza y covarianza de una cartera. Además, propuso la noción de la "frontera eficiente", que representa las combinaciones óptimas de activos que ofrecen el mayor rendimiento para un nivel de riesgo o la menor varianza para un nivel de rendimiento.

La literatura sobre la minimización de la varianza en la construcción de portafolios es extensa y ha sido un tema fundamental en la teoría financiera moderna. Varios factores pueden explicar por qué. En primer lugar, y de acuerdo con Coqueret, G. (2015) las carteras de varianza mínima no se ven afectadas por errores de estimación de los rendimientos esperados, que se sabe que tienen un impacto considerable en el desempeño fuera de la muestra de las estrategias de índice máximo de Sharpe, tal cual lo demuestra (Britten, M. 1999); (Merton, R. C. 1980) y (Kondor, I. et al., 2007).

En segundo lugar, se ha verificado por Haugen, R., y Baker, N. (1991) y Baker, M. et al., (2011) que sistemáticamente las acciones de baja volatilidad no obtienen peores resultados que sus homólogos de alta volatilidad, es decir que las acciones de baja volatilidad tienden a tener rendimientos que igualan o superan al mercado (Clarke, R. *et al.,* 2011).

Ha habido números avances interesantes considerando el rendimiento y el riesgo, por ejemplo, Zhai, J., y Bai, M. (2018) desarrollan un modelo de media-varianza para la optimización de carteras considerando el riesgo de fondo, la liquidez y el costo de transacción basado en la teoría de la incertidumbre, además analizan la característica de la frontera de la cartera considerando el riesgo de fondo independientemente aditivo, así mismo Haugen, R., y Baker, N., (1991) emplearon la teoría de la incertidumbre para estudiar un problema de selección de cartera incierta con riesgo de fondo.. Por otro lado, Ordoñez, J. (2004) y Lwin, K. T. et al., (2017) abordan la optimización de carteras en el marco del (VaR) medio, utilizando un modelo alternativo de media-varianza de Markowitz.

En el contexto latinoamericano, investigaciones recientes como las Amorocho, A. (2023) en el cual se investiga la presencia de la anomalía de rendimientos superiores en carteras de mínima varianza en comparación con las carteras ponderadas por capitalización de mercado en el contexto del mercado colombiano. De la torre, O., y Torre, M. (2017) propone el uso del portafolio de mínima varianza como método de ponderación para un benchmark de estrategia que evalúe el rendimiento de los fondos de pensiones en México. Para finalizar esta breve revisión, Zapata, C. (2022) propone un enfoque para la selección de portafolios socialmente responsables al integrar los criterios ambientales, sociales y de buen gobierno (ASG) en el modelo media-varianza (MV) de Markowitz.

En este estudio, se realiza un exhaustivo análisis de las carteras de varianza mínima, con un enfoque particular en la composición de estas carteras y en la forma en que se determinan los parámetros ponderados de los valores individuales. Nos centramos en acciones con un alto volumen de transacción en la Bolsa de Valores de Colombia (BVC), prestando especial atención al período durante la pandemia.

El artículo presenta una explicación detallada de los procedimientos cuantitativos esenciales utilizados para construir estas carteras de varianza mínima. Comenzamos en la sección 2, donde se detalla meticulosamente la metodología empleada, así como la fuente de datos utilizada para llevar a cabo este análisis.

En la sección 3, se describen los resultados obtenidos a partir de este análisis, destacando en un apartado la síntesis de los principales indicadores en la composición de las carteras de varianza mínima durante el período de estudio, y, por último, en la sección 4, se presentan las conclusiones derivadas de este estudio, donde se resumen los hallazgos clave.

## 2. METODOLOGÍA Y DATA

Este documento presenta un enfoque cuantitativo y descriptivo en su estructura. Adopta un diseño de estudio no experimental de naturaleza longitudinal para datos de panel. La recopilación de datos se lleva a cabo a través de la revisión documental centrada principalmente en fuentes de datos especializadas, en particular, Portal Financiero (Grupo AVAL).

Este análisis se fundamenta en los datos diarios de las empresas con mayor transacción para la fecha del 17 enero del 2024 en Colombia y abarca un período de tiempo que comprende desde 2015 hasta 2023. Se escogió el índice ICOLCAP como el $\mu_m$ dado que es el fondo de acciones colombianas compuesta por las 20 empresas más líquidas del mercado (BlackRock, 2024), y la tasa libre de riesgo $r_f$ es el TFIT16240724 (TES 2024), que por ser un bono soberano del gobierno colombiano garantiza un rendimiento, pero con menor riesgo.

Este estudio se divide en 5 submuestras (2015 – 2019; 2016 – 2020; 2020; 2020 – 2023; 2023) con el propósito de examinar el comportamiento de los portafolios en tres contextos distintos: uno caracterizado por la estabilidad, otro por la reacción ante la pandemia, y el tercero en el período posterior a la pandemia.

**Tabla 1.** Activos con mayor volumen de transacción.

| Nemotécnico | Volumen | Cantidad | Emisor |
|---|---|---|---|
| PFBCOLOMBIA | 13,833,696,500.00 | 445,398.00 | BANCO BANCOLOMBIA |
| ECOPETROL | 11,786,248,575.00 | 5,092,743.00 | ECOPETROL |
| ISA | 9,820,599,780.00 | 554,381.00 | ISA INTERCONEXION ELECTRICA |
| BCOLOMBIA | 8,665,346,060.00 | 247,677.00 | BANCO BANCOLOMBIA |

Fuente: Elaboración propia con datos del Grupo Aval (2024)

**Conceptos y Formulación matemática**

Para iniciar el cálculo del portafolio de Markowitz se inicia en primer lugar con el cálculo de los retornos $\mu$ de los activos, donde $P_i$ es el precio del activo en el presente y $P_{i-1}$ es el precio del mismo activo en el periodo inmediatamente anterior.

$$\mu_i = \frac{P_i}{P_{i-1}} - 1; \ (i = 1, \dots, n)$$

(1)

Posteriormente se multiplica Π los factores obtenidos en el paso anterior, se obtiene el producto de todos los incrementos (o decrecimientos) por periodo. Para adecuar los valores al contexto de tasas de crecimiento, se realiza la operación de sumar 1 a cada valor en el rango. Para que las tasas de crecimiento puedan ser factores multiplicativos.

$$\tau = \left(1 + \prod_{i=1}^{n=n+1} \mu_i \right)^{\frac{1}{n}} - 1$$

(2)

Posteriormente, se eleva el producto resultante a la potencia de $1/n$, lo que equivale a calcular la raíz $n$ del producto. La elección de la raíz $n$ se justifica por la división del conjunto total en $n$ periodos, indicando así el cálculo de la tasa de rendimiento promedio geométrico anualizada. Finalmente, restar 1 al resultado final ajusta la cifra obtenida, proporcionando la tasa de rendimiento promedio geométrico anualizada.

Para determinar la volatilidad anualizada, se procede a calcular la desviación estándar del conjunto de datos diarios. Posteriormente, se ajusta este resultado considerando el número de días hábiles en un año, que comúnmente se establece en 252 días bursátiles.

$$\sigma_i = \sqrt{\frac{\sum_{i=1}^{n=n+1}\left(\mu_i - \frac{\mu_i}{n}\right)^2}{n-1}} * 252$$

(3)

La beta se utiliza para calcular el riesgo sistemático o riesgo no diversificable de un activo dentro de una cartera. El riesgo sistemático es aquel que no puede ser eliminado a través de la diversificación, ya que está relacionado con los movimientos generales del mercado. En términos más sencillos, la beta se calcula tomando la covarianza entre los rendimientos del activo y del mercado, y dividiendo esto por la varianza del rendimiento del mercado.

$$\beta = \frac{\frac{\sum_{i=1}^{n=n+1}\left(\mu_i - \frac{\mu_i}{n}\right)\left(m_i - \frac{m_i}{n}\right)}{n}}{\frac{\sum_{i=1}^{n=2080}\left(\mu_i - \frac{\mu_i}{n}\right)^2}{n-1}}$$

(4)

- Si el valor de $\beta = 1$, indica que el activo tiene una neutralidad en su tasa de retorno, es decir, sus movimientos al alza y a la baja serán proporcionales a los de la cartera de mercado.
- En cambio, si $\beta > 1$, denota que el activo presenta un mayor nivel de riesgo, lo que implica que sus fluctuaciones, ya sean al alza o a la baja, serán más pronunciadas en comparación con el promedio de la cartera del mercado.
- Si $\beta < 1$, señala que el activo tiene un menor nivel de riesgo, lo que implica que sus variaciones, tanto al alza como a la baja, serán menos significativas que el promedio de la cartera del mercado.

Los valores negativos de $-\beta$ se aplican a activos cuyos rendimientos se mueven en dirección opuesta a los del mercado. En otras palabras, cuando el rendimiento del mercado aumenta, el rendimiento de las acciones de la empresa disminuye (Sharpe, 1964; Alqisie y Alqurran, 2016).

Uno de los requisitos para poder resolver el problema de optimización que supone encontrar el portafolio de mínima varianza es tener los retornos esperados, para determinar las expectativas de retorno $E(R)$ se utiliza el modelo de Valoración de Activos Financieros (CAPM), este modelo proporciona una estimación de la tasa de rendimiento esperada de un activo en función de su beta, $\beta$, la tasa libre de riesgo $r_f$ y el rendimiento del mercado $\mu_m$.

$$E(R) = r_f + \beta(\mu_m - r_f)$$

(5)

La función $\beta(\mu_m - r_f)$ representa el premio por riesgo de mercado. La beta ($\beta$) cuantifica la sensibilidad del activo respecto a los movimientos del mercado. La tasa de rendimiento libre de riesgo $(r_f)$ es la tasa de retorno que se obtendría sin asumir riesgo, por ejemplo, a través de inversiones en bonos del gobierno. El $(\mu_m - r_f)$ representa el premio por riesgo de mercado, que es la compensación adicional que los inversores requieren por asumir el riesgo inherente al mercado de acciones en lugar de invertir en activos sin riesgo (Gallardo, J. 2018).

La matriz de covarianza proporciona información sobre cómo los rendimientos de dos activos se mueven juntos. Si la covarianza es positiva, significa que los rendimientos de los dos activos tienden a moverse en la misma dirección. Si es negativa, los rendimientos tienden a moverse en direcciones opuestas. La diagonal principal de la matriz contiene las varianzas individuales de cada activo.

$$A = \begin{bmatrix} Var(\mu_i) & Cov(\mu_i, \mu_2) & \cdots & Cov(\mu_i, \mu_n) \\ Cov(\mu_2, \mu_i) & Var(\mu_2) & \cdots & Cov(\mu_2, \mu_n) \\ \vdots & \vdots & \ddots & \vdots \\ Cov(\mu_n, \mu_i) & Cov(\mu_n, \mu_2) & \cdots & Var(\mu_n) \end{bmatrix}$$

(6)

Esta inversa se utiliza para examinar la relación entre los rendimientos de los activos en el portafolio. Cuando un elemento de la matriz de covarianza invertida es positivo, indica que los rendimientos de los activos tienden a moverse en la misma dirección. Por el contrario, si el elemento es negativo, sugiere que los rendimientos de los activos tienen una tendencia a moverse en direcciones opuestas.

$$A^{-1}$$

(7)

Una vez hallada la matriz inversa de la covarianza este se multiplica con una matriz de vectores $1(n)$ haciendo referencia a los $n$ activos del portafolio (Clarke, R. *et al.,* 2011) resultado de este producto una matriz $h$ de la misma dimensión. Este vector refleja la varianza y covarianza total de una cartera igualmente ponderada.

$$h = \begin{bmatrix} 1 & 1 & 1 & 1 \end{bmatrix}[A^{-1}]$$

(8)

Después de calcular las expectativas de retorno, se procede a realizar la multiplicación de estas por la matriz inversa de covarianza, generando así el vector $g$. Esta matriz resultante representa la contribución de cada activo al rendimiento esperado de la cartera, teniendo en cuenta tanto las expectativas de rendimiento como las covarianzas entre los activos.

$$g = [E(R)^{ac1} \quad E(R)^{ac2} \quad E(R)^{ac3} \quad E(R)^{ac4}][A^{-1}]$$

(9)

$$\alpha = [1 \quad 1 \quad 1 \quad 1][h^t]$$

(10)

La constate $\alpha$ representa una constante en términos de rendimiento, esta [constante] de acuerdo con Merton, R. C. (1972) se define como el producto escalar del vector con valores de 1 y $h$ (resultado de multiplicar la matriz de covarianza inversa $A^{-1}$ por el vector de unos). En términos castizos, representa una medida de la relación entre el vector de unos (que refleja la diversificación) y la matriz de covarianza inversa (que mide la variabilidad entre los rendimientos de activos).

$$B = [1 \quad 1 \quad 1 \quad 1][g^t]$$

(11)

Similar a $\alpha$, pero en lugar de $h$, se emplea el vector transpuesto resultante del producto entre el vector de unos y la matriz inversa $h$. Esta elección refleja la relación entre la diversificación, representada por la matriz (1x4) con valores 1, y los rendimientos reales de los activos.

$$\gamma = [E(R)^{ac1} \quad E(R)^{ac2} \quad E(R)^{ac3} \quad E(R)^{ac4}][g^t]$$

(12)

Al igual que $B$, pero excluyendo el vector de unos, en su lugar se utiliza el CAPM (vector de rendimientos) junto con la transpuesta de $[g^t]$. Esta constante refleja la relación cuadrática entre los rendimientos y la matriz de covarianza inversa, ya que se multiplican los rendimientos por sí mismos.

$$\delta = \alpha(\gamma) - B^2$$

(13)

Aquí, $\delta$ es un parámetro de escala que combina las relaciones cuadráticas y lineales entre los rendimientos y las covarianzas. Es de notar que $\delta$ tiene una naturaleza cuadrática, ya que implica restar un término cuadrático $B^2$ de otro término cuadrático $(\alpha.\gamma)$.

El modelo teórico de Markowitz (1952) postula que los inversionistas organizan sus carteras incluyendo activos riesgosos. Sin embargo, el teorema de separación de Tobin (1958) expande la propuesta original de Markowitz al introducir un activo libre de riesgo ($r_f$) al conjunto inicial de activos riesgosos. En consecuencia, el portafolio óptimo se sitúa en el punto de tangencia que se forma entre la Línea de Mercado de Capitales (LMC) y la frontera eficiente de Markowitz, un punto conocido como el portafolio de mercado (Calderón, J. 2016).

Para determinar la línea de mercado de capitales esto implica la identificación crucial del punto de retorno tangente $R_t$. Este punto representa el encuentro entre la tasa de rendimiento libre de riesgo y la curva de frontera eficiente en el eje $y$ del gráfico. De manera análoga, el punto de riesgo tangente $\sigma_{Rt}$ se posiciona en el eje $x$, marcando el punto de intersección entre la línea de frontera eficiente y la línea de mercado de capitales

$$R_t = \frac{(\gamma - B)r_f}{(B - \alpha)r_f}$$

(14)

$$\sigma_{Rt} = \sqrt{\frac{[\alpha(R_t{}^2) - 2B(R_t) + \gamma]}{\delta}}$$

(15)

La LMC de acuerdo con la definición de Echeverri, A., y Giraldo, A. (2008) es una línea recta que conecta la tasa libre de riesgo con la cartera de mercado y refleja la relación entre riesgo y rendimiento y se calcula como la diferencia entre la tasa de rendimiento de la cartera de mercado y la tasa libre de riesgo, dividida por la volatilidad (o riesgo) de la cartera de mercado.

$$p_{lmc} = \frac{(R_t - r_f)}{\sigma_{Rt}}$$

(16)

Ahora procedemos a calcular el riesgo de la frontera eficiente cuando se quiere un retorno determinado de $x$ unidad. La Frontera Eficiente (FE) representa el conjunto óptimo de portafolios que logran la mejor combinación entre riesgo y retorno, considerando los activos disponibles en el mercado. De acuerdo con Morato Leyton, H. (2023, p. 19) cuando los activos se sitúan por encima de la frontera, indica la inexistencia de portafolios factibles más allá de esta[1].

$$Riesgo = \sqrt{\frac{[\alpha(x^2) - 2B(x) + \gamma]}{\delta}}; \quad x = 1,2,3\dots n$$

(17)

Y tambien se procede a calcular la línea de la LMC cuando se tiene un riesgo determinado $v$ unidad.

$$LMC = r_f + v(p_{lmc}); \quad v = 1,2,3\dots n$$

(18)

Una vez identificado el valor de la mínima varianza con su respectivo retorno se establecen los pesos $\omega$, para lo cual se siguen estos pasos:

$$\lambda = \frac{\gamma - B(\mu^*)}{\delta}$$

(19)

$$\theta = \frac{\alpha(\mu^* - B)}{\gamma}$$

(20)

$$\omega = \lambda[h] + \theta[g]$$

(21)

---

[1] En otras palabras, no se pueden encontrar combinaciones de activos que ofrezcan un rendimiento superior para un nivel de riesgo dado o un riesgo inferior para un nivel de rendimiento dado.

La validación del rendimiento del portafolio se logra al multiplicar las expectativas de retorno de cada activo por sus respectivos pesos.

$$n = [E(R)][\omega]$$

(22)

Para hallar la varianza $v$ del portafolio, es necesario multiplicar la matriz transpuesta $\omega^t$ de los pesos por la matriz de covarianzas $A$ y luego por la matriz de pesos $\omega$.

$$v = \omega^t(A)\omega$$

(23)

La $v$ representa la dispersión de los rendimientos de los activos dentro del portafolio. Sin embargo, para expresar este riesgo de manera más intuitiva y en la misma escala que los rendimientos, se utiliza la raíz cuadrada de la varianza $\sqrt{v}$, que se conoce como la desviación estándar.

$$riesgo = \sqrt{v}$$

(24)

Finalmente, procederemos a calcular el índice de Sharpe para evaluar la rentabilidad ajustada al riesgo en nuestro portafolio. El índice de Sharpe es una métrica que, de acuerdo con Contreras, O. E. *et al.,* (2015) permite medir la eficiencia de una inversión al considerar el rendimiento del portafolio $R_p$ en relación con el riesgo asumido.

$$Sh = \frac{R_p - r_f}{\sigma_i^2}$$

(25)

## 3. RESULTADOS

**Figura 1.** Comportamiento histórico periodo 2015 - 2023.

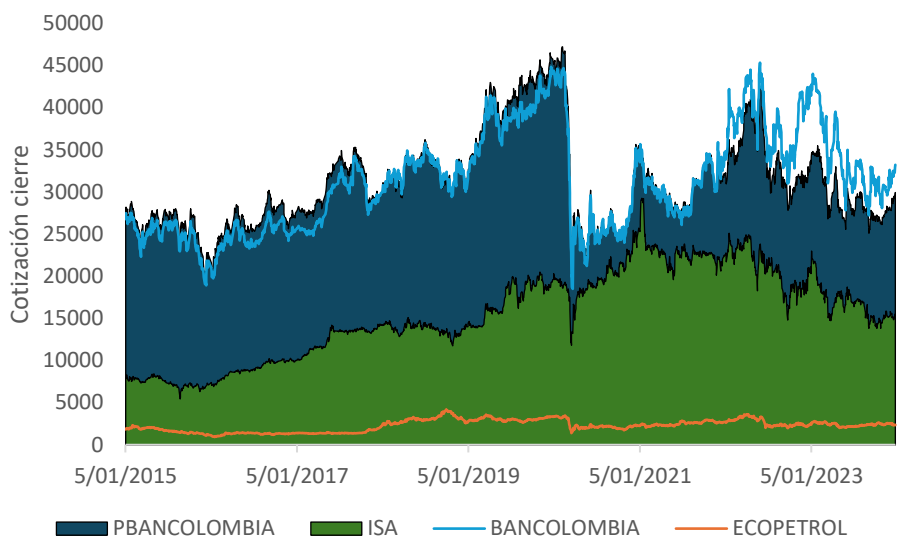

Fuente: Elaboración propia (2024)

**Tabla 2.** Indicadores anuales periodo 2015 - 2023.

| Concepto | PBANCOLOMBIA | ECOPETROL | ISA | BANCOLOMBIA | ICOLCAP |
|---|---|---|---|---|---|
| Rendimiento | 0,81% | 2,86% | 7,72% | 2,21% | -1,88% |
| Volatilidad | 29,56% | 38,65% | 34,23% | 34,69% | 20,67% |
| Beta | 43,85% | 29,19% | 24,49% | 33,75% | 100,00% |
| CAMP | 3,03% | 4,32% | 4,73% | 3,92% | -1,88% |
| Sharpe | -20,50% | -10,37% | 2,47% | -13,43% | -42,33% |
| Traynor | -13,82% | -13,73% | 3,45% | -13,80% | -8,75% |

Fuente: Elaboración propia (2024)

En relación con la tabla anterior, el activo de ISA destaca como la opción más favorable para inversores en busca de rendimientos robustos, respaldado por sólidos indicadores como CAMP, Sharpe y Traynor. Su desempeño superior la posiciona como una elección atractiva para aquellos dispuestos a asumir mayores riesgos en función de mayores retornos.

En contraste, PBANCOLOMBIA es una alternativa para inversores que priorizan la minimización del riesgo, ofreciendo un rendimiento más moderado. ECOPETROL, con su rendimiento moderado y alta volatilidad, se presenta como una opción equilibrada para aquellos que buscan un balance entre riesgo y retorno. Similarmente, BANCOLOMBIA ofrece un rendimiento y volatilidad moderados, siendo una opción atractiva para inversores que buscan mantener un equilibrio prudente entre riesgo y retorno en sus inversiones

**Tabla 3.** Matriz de covarianzas periodo 2015 - 2023.

| Covarianza | PBANCOLOMBIA | ECOPETROL | ISA | BANCOLOMBIA |
|---|---|---|---|---|
| PBANCOLOMBIA | 8,74% | 4,92% | 3,73% | 7,82% |
| ECOPETROL | 4,92% | 14,94% | 3,99% | 4,78% |
| ISA | 3,73% | 3,99% | 11,72% | 4,68% |
| BANCOLOMBIA | 7,82% | 4,78% | 4,68% | 12,03% |

Fuente: Elaboración propia (2024)

En el análisis de covarianzas entre activos financieros, se observa que la relación positiva entre PBANCOLOMBIA y ECOPETROL es del 4.92%, indicando movimientos sincronizados. De manera similar, la covarianza positiva de 3.99% entre ECOPETROL e ISA sugiere una tendencia a movimientos conjuntos. Asimismo, la relación positiva de 3.73% entre PBANCOLOMBIA e ISA señala una asociación en sus movimientos. Finalmente, la covarianza positiva de 4.68% entre BANCOLOMBIA e ISA indica una relación de movimiento conjunto.

**Tabla 4.** Matriz inversa de covarianzas periodo 2015 - 2023.

| C. Inversa | PBANCOLOMBIA | ECOPETROL | ISA | BANCOLOMBIA |
|---|---|---|---|---|
| PBANCOLOMBIA | 2955% | -392% | -120% | -1718% |
| ECOPETROL | -392% | 847% | -155% | -22% |
| ISA | -120% | -155% | 1052% | -269% |
| BANCOLOMBIA | -1718% | -22% | -269% | 2062% |

Fuente: Elaboración propia (2024)

Dentro del grupo, se evidencia que PBANCOLOMBIA representa la acción más riesgosa, aunque ofrece la menor rentabilidad potencial. En contraste, ECOPETROL e ISA muestran una correlación relativamente baja con las demás acciones, lo que las convierte en opciones atractivas para la diversificación de un portafolio, ya que presentan menor riesgo en comparación con BANCOCOLOMBIA y ofrecen una mayor estabilidad en términos de rendimiento.

**Vector de varianza y covarianza igualmente ponderada.**

$$h = \begin{bmatrix} 1 & 1 & 1 & 1 \end{bmatrix} \begin{bmatrix} 2955\% & -392\% & -120\% & -17,18\% \\ -392\% & 847\% & -155\% & -22\% \\ -120\% & -155\% & 1052\% & -269\% \\ -1718\% & -22\% & -269\% & 2062\% \end{bmatrix} = \begin{bmatrix} 724\% & 279\% & 507\% & 52\% \end{bmatrix}$$

(1)

Bajo la ponderación proporcionada por las matrices previas, la combinación lineal resultante $h$ de las covarianzas inversas indica una contribución positiva al riesgo total en la cartera. Esto implica que, según la estructura de las covarianzas inversas y sus ponderaciones, los activos PBANCOLOMBIA, ECOPETROL, ISA y BANCOLOMBIA tienen una influencia conjunta en el riesgo total de la cartera.

$$g = \begin{bmatrix} 3{,}03\% & 4{,}32\% & 4{,}73\% & 3{,}92\% \end{bmatrix} \begin{bmatrix} 2955\% & -392\% & -120\% & -17{,}18\% \\ -392\% & 847\% & -155\% & -22\% \\ -120\% & -155\% & 1052\% & -269\% \\ -1718\% & -22\% & -269\% & 2062\% \end{bmatrix}$$

$$= \begin{bmatrix} -0{,}28\% & 16{,}5\% & 28{,}8\% & 14{,}9\% \end{bmatrix}$$

(2)

Se observa una ligera disminución del rendimiento esperado en relación con el activo PBANCOLOMBIA, con un valor de -0.28%. Contrariamente, se evidencia un notable aumento en los rendimientos asociados con los activos ECOPETROL, ISA y BANCOLOMBIA, registrando valores de 16.5%, 28.8% y 14.9%, respectivamente.

$$A = \begin{bmatrix} 1 & 1 & 1 & 1 \end{bmatrix} \begin{bmatrix} 724\% \\ 279\% \\ 507\% \\ 52\% \end{bmatrix} = 1562\%$$

(3)

$$B = \begin{bmatrix} 1 & 1 & 1 & 1 \end{bmatrix} \begin{bmatrix} -0{,}28\% \\ 16{,}5\% \\ 28{,}8\% \\ 14{,}9\% \end{bmatrix} = 60{,}03\%$$

(4)

$$\gamma = \begin{bmatrix} 3{,}03\% & 4{,}32\% & 4{,}73\% & 3{,}92\% \end{bmatrix} \begin{bmatrix} -0{,}28\% \\ 16{,}5\% \\ 28{,}8\% \\ 14{,}9\% \end{bmatrix} = 2{,}65\%$$

(5)

$$\delta = 1562\%(2{,}65\%) - 60{,}03\%^2 = 5{,}43\%$$

(6)

$$R_t = \frac{(2{,}65\% - 60{,}03\%)6{,}87\%}{(60{,}03\% - 1562\%)6{,}87\%} = 3{,}11\%$$

(7)

En este punto 3,11% representa el punto de encuentro entre la tasa de rendimiento libre de riesgo y la curva de frontera eficiente en el eje y del gráfico.

$$\sigma_{Rt} = \sqrt{\frac{[1562\%(3{,}11\%^2) - 2(60{,}03\%)(3{,}11\%) + 2{,}65\%]}{5{,}43\%}} = 28{,}2\%$$

(8)

Por otro lado, 28,2% representa el punto de encuentro entre la tasa de rendimiento libre de riesgo y la curva de frontera eficiente en el eje y del gráfico, representando la cartera óptima que equilibra riesgo y rendimiento.

$$p_{lmc} = \frac{(3{,}11\% - 6{,}87\%)}{28{,}2\%} = -13{,}3\%$$

(9)

En este caso, la pendiente es negativa, lo que indica que la cartera de mercado ofrece una tasa de rendimiento esperada menor en comparación con la tasa libre de riesgo. Esta situación sugiere que la cartera de mercado no está proporcionando una compensación suficiente por el riesgo asumido. Ahora se procede a calcular la mínima varianza cuando se quiere un retorno de terminado $x$.

$$Riesgo = \sqrt{\frac{[15{,}62(0{,}145^2) - 2(2{,}26)(0{,}145) + 0{,}35]}{0{,}442}} = 25{,}3\%$$

(10)

Ahora se procede a calcular la LMC cuando se tiene un riesgo determinado $v$ unidad, como el proceso requiere de una cantidad amplia de datos.

$$LMC = 6{,}87\% + v(-13{,}3\%); \quad v = 1,2,3 \ldots n$$

(11)

Una vez identificado el valor con mínima varianza (25,3%) para el retorno (3,8%), se establecen los pesos del portafolio, para lo cual es necesario el caculo de $\lambda$ y $\theta$, los cuales son los coeficientes que multiplican al vector $h$ y $g$.

$$\lambda = \frac{2{,}65\% - 60{,}03\%(3{,}8\%)}{5{,}43\%} = 6{,}9\%$$

(12)

$$\theta = \frac{1562\%(3{,}8\% - 60{,}03\%)}{5{,}43\%} = -12{,}3\%$$

(13)

**Pesos del portafolio**

$$\omega = 6{,}9\%[724\% \quad 279\% \quad 507\% \quad 52\%] + -12{,}3\%[-0{,}28\% \quad 16{,}5\% \quad 28{,}8\% \quad 14{,}9\%]$$

(14)

**Tabla 5.** Cálculo de los pesos a partir de la ecuación #.

| Nemotécnico | Formula | Peso |
|---|---|---|
| PBANCOLOMBIA | $61\%(724\%) + -12{,}3\%(-0{,}28\%)$ | 49,8% |
| ECOPETROL | $61\%(279\%) + -12{,}3\%(16{,}5\%)$ | 17,1% |
| ISA | $61\%(507\%) + -12{,}3\%(28{,}8\%)$ | 31,3% |
| BANCOLOMBIA | $61\%(52\%) + -12{,}3\%(14{,}9\%)$ | 1,8% |
| Total | | 100% |

Fuente: Elaboración propia (2024)

**Rendimiento del portafolio**

El retorno del portafolio se puede reconfirmar con la suma producto de $\omega$ y $E(R)$ lo siguiente:

$$n = \begin{bmatrix} 49{,}8\% \\ 17{,}1\% \\ 31{,}3\% \\ 1{,}8\% \end{bmatrix} \begin{bmatrix} 3{,}03\% \\ 4{,}32\% \\ 4{,}73\% \\ 3{,}92\% \end{bmatrix} = \begin{bmatrix} 1{,}51\% \\ 0{,}74\% \\ 1{,}48\% \\ 0{,}07\% \end{bmatrix}$$

(15)

$$R_{p2} = \sum_{i=1}^{n=4} n = 3{,}8\%$$

(16)

El portafolio registró un retorno anual del 3.8% durante el período examinado. Destaca la significativa participación de PBANCOLOMBIA, con un peso del 49.8%, seguido por ISA con un 31.3%

$$v = \begin{bmatrix} 49{,}8\% & 17{,}1\% & 31{,}3\% & 1{,}8\% \end{bmatrix} \begin{bmatrix} 8{,}74\% & 4{,}92\% & 3{,}73\% & 7{,}82\% \\ 4{,}92\% & 14{,}94\% & 3{,}99\% & 4{,}78\% \\ 3{,}73\% & 3{,}99\% & 11{,}72\% & 4{,}68\% \\ 7{,}82\% & 4{,}78\% & 4{,}68\% & 12\% \end{bmatrix} \begin{bmatrix} 49{,}8\% \\ 17{,}1\% \\ 31{,}3\% \\ 1{,}8\% \end{bmatrix} = 6{,}40\%$$

(17)

$$\sigma_i^2 = \sqrt{6{,}40\%} = 25{,}3\%$$

(18)

$$Sh = \frac{3{,}8\% - 6{,}87\%}{25{,}3\%} = -12{,}3\%$$

(19)

Durante el periodo 2015-2023, la varianza $v$ del portafolio fue del 6,40%, indicando la dispersión de rendimientos. El riesgo anual $\sigma_i^2$ asociado al portafolio fue del 25,3%, señalando la volatilidad de los resultados de inversión. Sin embargo, el índice de Sharpe $Sh$, que evalúa la rentabilidad ajustada al riesgo, mostró un preocupante -12,3%. Este resultado indica que el rendimiento del activo es inferior al riesgo que se está asumiendo al invertir en el mismo.

**Figura 2.** Frontera eficiente y LMC en el periodo 2015 - 2023.

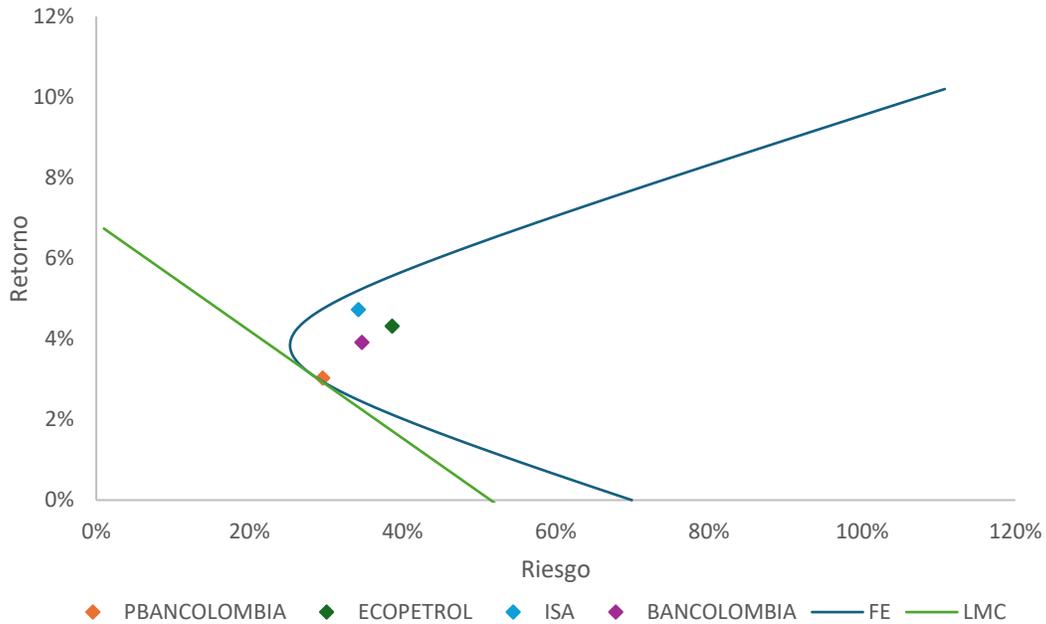

Fuente: Elaboración propia (2024)

Al analizar la figura anterior, el activo de PBANCOCOLOMBIA cumple con los requisitos de la Frontera Eficiente (FE) al ubicarse en la línea óptima que ofrece el mejor equilibrio entre riesgo y retorno. Esto significa que han logrado la eficiencia al no haber otros activos que superen su rendimiento para el mismo nivel de riesgo o que ofrezcan un riesgo inferior para el mismo nivel de rendimiento.

**Figura 3.** Comportamiento histórico periodo 2015 - 2019.

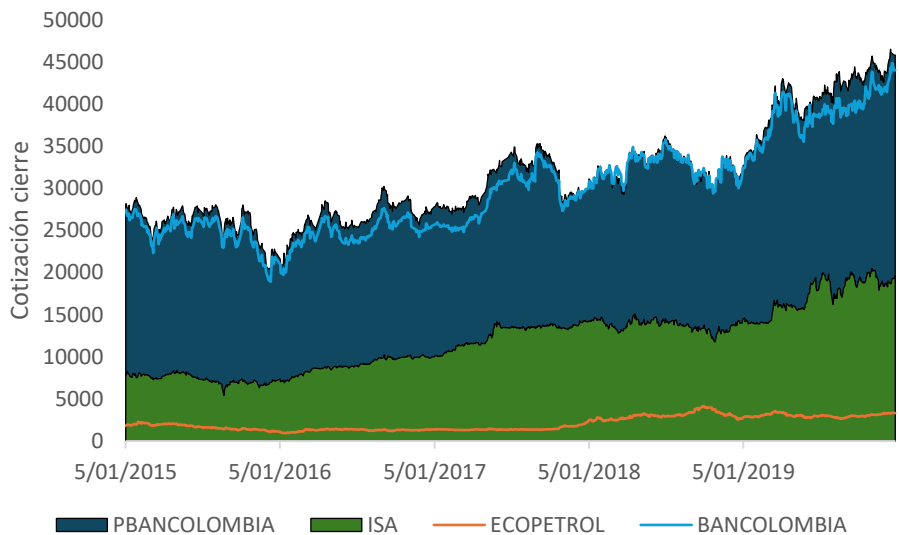

Fuente: Elaboración propia (2024)

*Después de analizar los resultados en la primera sección, en esta y las siguientes partes se presentan los hallazgos numéricos por sí mismos, dejando la interpretación al lector en función de los resultados iniciales.*

**Tabla 6.** Indicadores periodo 2015 - 2019.

| Concepto | PBANCOLOMBIA | ECOPETROL | ISA | BANCOLOMBIA | ICOLCAP |
|---|---|---|---|---|---|
| Rendimiento | 10,48% | 12,80% | 19,84% | 10,05% | 3,20% |
| Volatilidad | 23,56% | 33,84% | 25,43% | 25,75% | 14,40% |
| Beta | 33,55% | 21,89% | 19,61% | 30,21% | 100,00% |
| CAMP | 5,43% | 5,82% | 5,89% | 5,54% | 3,20% |
| Sharpe | 16,71% | 18,48% | 52,26% | 13,59% | -23,24% |
| Traynor | 11,73% | 28,58% | 67,79% | 11,59% | -3,35% |

Fuente: Elaboración propia (2024).

**Tabla 7.** Matriz de covarianzas periodo 2015 - 2019.

| Covarianza | PBANCOLOMBIA | ECOPETROL | ISA | BANCOLOMBIA |
|---|---|---|---|---|
| PBANCOLOMBIA | 5,55% | 2,53% | 1,22% | 5,02% |
| ECOPETROL | 2,53% | 11,45% | 1,62% | 2,50% |
| ISA | 1,22% | 1,62% | 6,47% | 1,47% |
| BANCOLOMBIA | 5,02% | 2,50% | 1,47% | 6,63% |

Fuente: Elaboración propia (2024).

**Tabla 8.** Matriz inversa de covarianzas periodo 2015 - 2019.

| C. Inversa | PBANCOLOMBIA | ECOPETROL | ISA | BANCOLOMBIA |
|---|---|---|---|---|
| PBANCOLOMBIA | 5840% | -352% | -43% | -4279% |
| ECOPETROL | -352% | 990% | -165% | -70% |
| ISA | -43% | -165% | 1657% | -272% |
| BANCOLOMBIA | -4279% | -70% | -272% | 4833% |

Fuente: Elaboración propia (2024)

$$h = \begin{bmatrix} 1 & 1 & 1 & 1 \end{bmatrix} \begin{bmatrix} 5840\% & -352\% & -43\% & -4279\% \\ -352\% & 990\% & -165\% & -70\% \\ -43\% & -165\% & 1657\% & -272\% \\ -4279\% & -70\% & -272\% & 4833\% \end{bmatrix} = \begin{bmatrix} 1167\% & 402\% & 1177\% & 213\% \end{bmatrix}$$

(1)

$$g = \begin{bmatrix} 5,4\% & 5,8\% & 5,9\% & 5,5\% \end{bmatrix} \begin{bmatrix} 2955\% & -392\% & -120\% & -17,18\% \\ -392\% & 847\% & -155\% & -22\% \\ -120\% & -155\% & 1052\% & -269\% \\ -1718\% & -22\% & -269\% & 2062\% \end{bmatrix}$$
$$= \begin{bmatrix} 56,9\% & 24,8\% & 70,6\% & 15,4\% \end{bmatrix}$$

(2)

$$A = \begin{bmatrix} 1 & 1 & 1 & 1 \end{bmatrix} \begin{bmatrix} 1167\% \\ 402\% \\ 1177\% \\ 213\% \end{bmatrix} = 2959\% \tag{3}$$

$$B = \begin{bmatrix} 1 & 1 & 1 & 1 \end{bmatrix} \begin{bmatrix} 56,9\% \\ 24,8\% \\ 70,6\% \\ 15,4\% \end{bmatrix} = 168\% \tag{4}$$

$$\gamma = \begin{bmatrix} 5,4\% & 5,8\% & 5,9\% & 5,5\% \end{bmatrix} \begin{bmatrix} 56,9\% \\ 24,8\% \\ 70,6\% \\ 15,4\% \end{bmatrix} = 9,6\% \tag{5}$$

$$\delta = 2959\%(9,65\%) - 167\%^2 = 0,88\% \tag{6}$$

$$R_t = \frac{(9,65\% - 168\%)6,55\%}{(168\% - 2959\%)6,55\%} = 5,56\% \tag{7}$$

$$\sigma_{Rt} = \sqrt{\frac{[2959\%(5,56\%^2) - 2(168\%)(5,56\%) + 9,5\%]}{0,88\%}} = 19,5\% \tag{8}$$

$$p_{lmc} = \frac{(3,11\% - 6,87\%)}{28,2\%} = -5,07\% \tag{9}$$

$$Riesgo = \sqrt{\frac{[2959\%(5,7\%^2) - 2(168\%)(5,7\%) + 9,5\%]}{0,88\%}} = 18,45\% \tag{10}$$

$$LMC = 6,55\% + v(-5,07\%); \quad v = 1,2,3 \dots n \tag{11}$$

$$\lambda = \frac{9,55\% - 167,8\%(5,7\%)}{0,88\%} = -1,87\% \tag{12}$$

$$\theta = \frac{2959\%(5,7\% - 167,8\%)}{0,88\%} = 92,6\% \tag{13}$$

$$\omega = -1,87\%[1167\% \quad 402\% \quad 1177\% \quad 213\%] + 92,6\%[56,9\% \quad 24,8\% \quad 70,6\% \quad 15,4\%] \tag{14}$$

**Tabla 9.** Cálculo de los pesos a partir de la ecuación 14.

| Nemotécnico | Formula | Peso |
|---|---|---|
| PBANCOLOMBIA | $-1,87\%(1167\%) + 92,6\%(56,9\%)$ | 30,8% |
| ECOPETROL | $-1,87\%(402\%) + 92,6\%(24,8\%)$ | 15,4% |
| ISA | $-1,87\%(1177\%) + 92,6\%(70,6\%)$ | 43,3% |
| BANCOLOMBIA | $-1,87\%(213\%) + 92,6\%(15,4\%)$ | 10,2% |
| | Total | 100% |

Fuente: Elaboración propia (2024)

El retorno del portafolio se puede reconfirmar con la suma producto de $\omega$ y $E(R)$ lo siguiente:

$$n = \begin{bmatrix} 30,8\% \\ 15,4\% \\ 43,3\% \\ 10,2\% \end{bmatrix} \begin{bmatrix} 5,43\% \\ 5,82\% \\ 5,89\% \\ 5,54\% \end{bmatrix} = \begin{bmatrix} 1,68\% \\ 0,90\% \\ 2,56\% \\ 057\% \end{bmatrix} \quad (15)$$

$$R_{p2} = \sum_{i=1}^{n=4} n = 5,7\% \quad (16)$$

$$v = [30,8\% \quad 15,4\% \quad 43,3\% \quad 10,2\%] \begin{bmatrix} 5,55\% & 2,53\% & 1,22\% & 5,02\% \\ 2,53\% & 11,4\% & 1,62\% & 2,5\% \\ 1,22\% & 1,62\% & 6,47\% & 1,47\% \\ 5,02\% & 2,5\% & 1,47\% & 6,63\% \end{bmatrix} \begin{bmatrix} 30,8\% \\ 15,4\% \\ 43,3\% \\ 10,2\% \end{bmatrix} = 3,41\% \quad (17)$$

$$\sigma_i^2 = \sqrt{3,41\%} = 18,41\% \quad (18)$$

$$Sh = \frac{5,7\% - 6,55\%}{18,45\%} = -4,6\% \quad (19)$$

**Figura 4.** Frontera eficiente y LMC en el periodo 2015 - 2019.

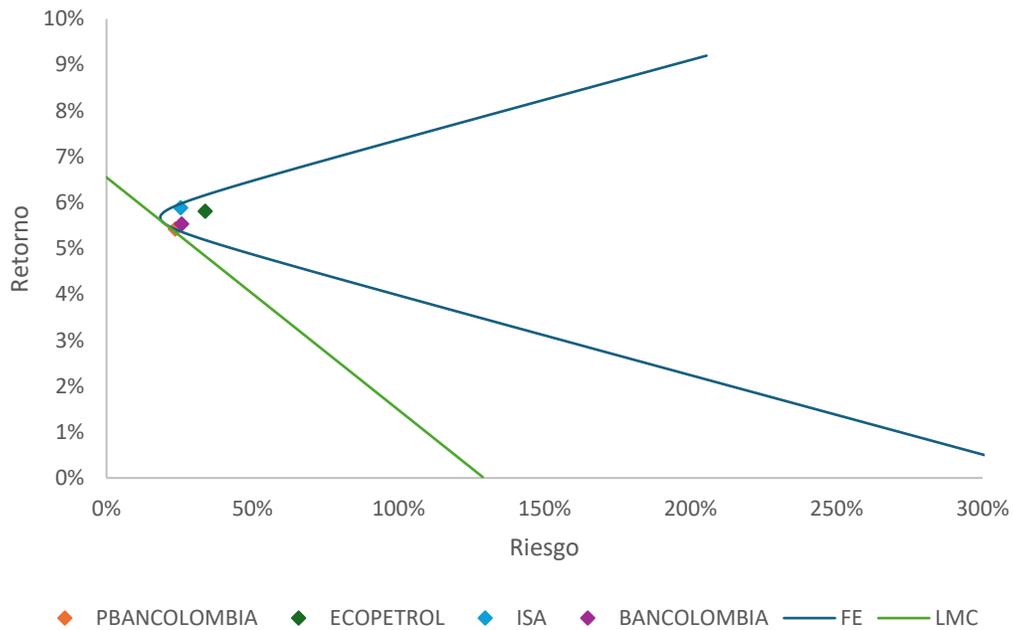

Fuente: Elaboración propia (2024)

**Figura 5.** Comportamiento histórico periodo 2016 - 2020.

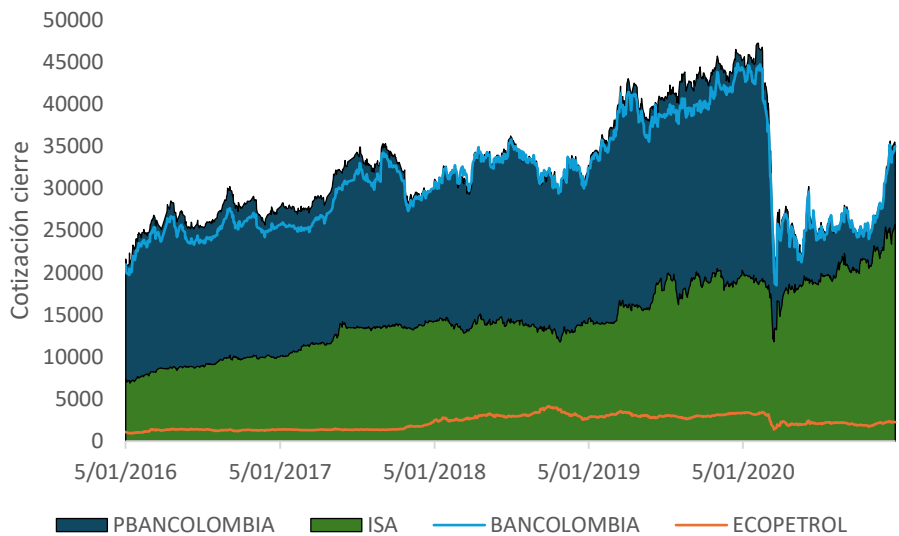

Fuente: Elaboración propia (2024)

**Tabla 10.** Indicadores financieros para el periodo 2016– 2020.

| Indicador | PBANCOLOMBIA | ECOPETROL | ISA | BANCOLOMBIA | ICOLCAP |
|---|---|---|---|---|---|
| Rendimiento | 10,20% | 15,13% | 28,34% | 10,77% | 4,53% |
| Volatilidad | 30,06% | 38,37% | 29,23% | 33,29% | 20,48% |
| Beta | 47,88% | 32,95% | 37,71% | 41,62% | 100,00% |
| CAMP | 5,27% | 5,48% | 5,41% | 5,35% | 4,53% |
| Sharpe | 14,17% | 23,95% | 76,65% | 14,50% | -6,86% |
| Traynor | 8,89% | 27,88% | 59,41% | 11,59% | -1,41% |

Fuente: Elaboración propia (2024)

**Tabla 11.** Matriz de covarianzas para el periodo 2016 – 2020.

| Covarianza | PBANCOLOMBIA | ECOPETROL | ISA | BANCOLOMBIA |
|---|---|---|---|---|
| PBANCOLOMBIA | 9,03% | 5,34% | 3,56% | 8,47% |
| ECOPETROL | 5,34% | 14,72% | 3,18% | 5,63% |
| ISA | 3,56% | 3,18% | 8,54% | 4,37% |
| BANCOLOMBIA | 8,47% | 5,63% | 4,37% | 11,08% |

Fuente: Elaboración propia (2024)

**Tabla 12.** Matriz inversa de covarianzas para el periodo 2016 – 2020.

| C. Inversa | PBANCOLOMBIA | ECOPETROL | ISA | BANCOLOMBIA |
|---|---|---|---|---|
| PBANCOLOMBIA | 4043% | -349% | -80% | -2880% |
| ECOPETROL | -349% | 883% | -114% | -137% |
| ISA | -80% | -114% | 1485% | -467% |
| BANCOLOMBIA | -2880% | -137% | -467% | 3356% |

Fuente: Elaboración propia (2024)

$$h = [1 \quad 1 \quad 1 \quad 1] \begin{bmatrix} 4043\% & -349\% & -80\% & -2880\% \\ -349\% & 883\% & -114\% & -137\% \\ -80\% & -114\% & 1485\% & -467\% \\ -2880\% & -137\% & -467\% & 3356\% \end{bmatrix} = [735\% \quad 283\% \quad 825\% \quad -128\%]$$

(1)

$$g = [5,2\% \quad 5,4\% \quad 5,4\% \quad 5,3\%] \begin{bmatrix} 2955\% & -392\% & -120\% & -17,18\% \\ -392\% & 847\% & -155\% & -22\% \\ -120\% & -155\% & 1052\% & -269\% \\ -1718\% & -22\% & -269\% & 2062\% \end{bmatrix}$$
$$= [35,3\% \quad 16,5\% \quad 44,9\% \quad -4,73\%]$$

(2)

$$A = [1 \quad 1 \quad 1 \quad 1] \begin{bmatrix} 735\% \\ 283\% \\ 825\% \\ -128\% \end{bmatrix} = 1715\%$$

(3)

$$B = [1 \quad 1 \quad 1 \quad 1] \begin{bmatrix} 35,3\% \\ 16,5\% \\ 44,9\% \\ -4,7\% \end{bmatrix} = 91,9\%$$

$$\gamma = [5{,}2\% \quad 5{,}4\% \quad 5{,}4\% \quad 5{,}3\%] \begin{bmatrix} 35{,}3\% \\ 16{,}5\% \\ 44{,}9\% \\ -4{,}7\% \end{bmatrix} = 4{,}94\% \tag{4}$$

$$\delta = 1715\%(4{,}94\%) - 91{,}9\%^2 = 0{,}10\% \tag{5}$$

$$R_t = \frac{(4{,}94\% - 91{,}9\%)5{,}94\%}{(91{,}9\% - 1715\%)5{,}94\%} = 5{,}31\% \tag{6}$$

$$\sigma_{Rt} = \sqrt{\frac{[1715\%(5{,}31\%^2)] - 2(91{,}9\%)(5{,}31\%) + 4{,}94\%]}{0{,}10\%}} = 25{,}3\% \tag{7}$$

$$p_{lmc} = \frac{(5{,}31\% - 5{,}94\%)}{25{,}3\%} = -2{,}5\% \tag{8}$$

$$Riesgo = \sqrt{\frac{[2959\%(5{,}4\%^2)] - 2(91{,}9\%)(5{,}4\%) + 4{,}9\%]}{0{,}10\%}} = 24{,}6\% \tag{9}$$

$$\lambda = \frac{4{,}94\% - 91{,}9\%(5{,}4\%)}{0{,}10\%} = -29{,}3\% \tag{10}$$

$$\theta = \frac{1715\%(5{,}4\% - 91{,}9\%)}{0{,}10\%} = 655\% \tag{11}$$

$$\omega = -29{,}3\%[735\% \quad 283\% \quad 825\% \quad -128\%] + 655\%[35{,}3\% \quad 16{,}5\% \quad 44{,}9\% \quad -4{,}73\%] \tag{12}$$

(13)

**Tabla 13.** Cálculo de los pesos a partir de la ecuación 13.

| Nemotécnico | Formula | Peso |
|---|---|---|
| PBANCOLOMBIA | $-29{,}3\%(735\%) + 655\%(35{,}3\%)$ | 16% |
| ECOPETROL | $-29{,}3\%(283\%) + 655\%(16{,}5\%)$ | 25% |
| ISA | $-29{,}3\%(825\%) + 655\%(44{,}9\%)$ | 53% |
| BANCOLOMBIA | $-29{,}3\%(-128\%) + 655\%(-4{,}73\%)$ | 6% |
| | Total | 100% |

Fuente: Elaboración propia (2024)

El retorno del portafolio se puede reconfirmar con la suma producto de $\omega$ y $E(R)$ lo siguiente:

$$n = \begin{bmatrix} 16\% \\ 25\% \\ 53\% \\ 6\% \end{bmatrix} \begin{bmatrix} 5,2\% \\ 5,4\% \\ 5,4\% \\ 5,3\% \end{bmatrix} = \begin{bmatrix} 0,84\% \\ 1,37\% \\ 2,84\% \\ 0,35\% \end{bmatrix}$$

(14)

$$R_{p2} = \sum_{i=1}^{n=4} n = 5,4\%$$

(15)

$$v = \begin{bmatrix} 16\% & 25\% & 53\% & 6\% \end{bmatrix} \begin{bmatrix} 9,03\% & 5,34\% & 3,56\% & 8,47\% \\ 5,34\% & 14,7\% & 3,18\% & 5,63\% \\ 3,56\% & 3,18\% & 8,54\% & 4,37\% \\ 8,47\% & 5,63\% & 4,37\% & 11\% \end{bmatrix} \begin{bmatrix} 16\% \\ 25\% \\ 53\% \\ 6\% \end{bmatrix} = 3,41\%$$

(16)

$$\sigma_i^2 = \sqrt{3,41\%} = 6,07\%$$

(17)

$$Sh = \frac{5,4\% - 5,94\%}{24,64\%} = -2,19\%$$

(18)

**Figura 6.** Frontera eficiente y LMC en el periodo 2016 - 2020.

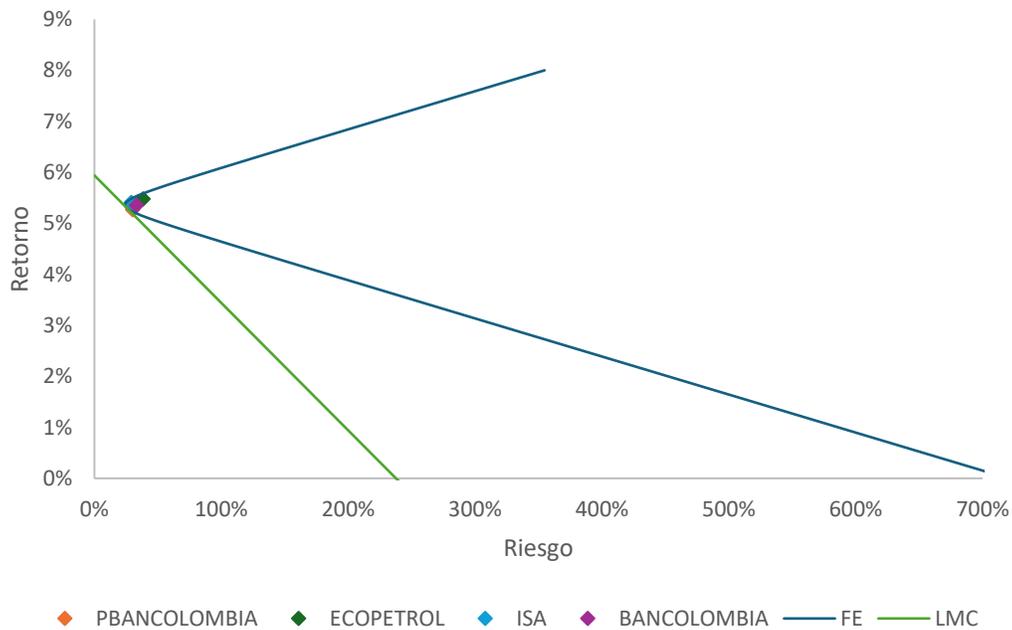

Fuente: Elaboración propia (2024)

**Figura 7.** Comportamiento periodo 2020.

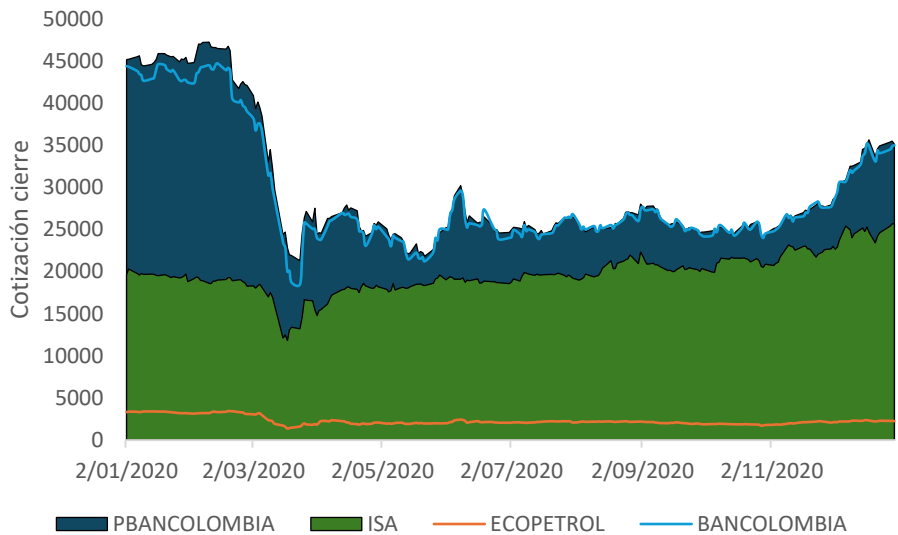

Fuente: Elaboración propia (2024)

**Tabla 14.** Indicadores financieros para el periodo 2020.

| Indicador | PBANCOLOMBIA | ECOPETROL | ISA | BANCOLOMBIA | ICOLCAP |
|---|---|---|---|---|---|
| Rendimiento | -21,38% | -31,05% | 30,75% | -21,41% | -12,91% |
| Volatilidad | 50,29% | 56,71% | 46,59% | 56,32% | 37,67% |
| Beta | 58,56% | 46,83% | 55,28% | 50,82% | 100,00% |
| CAMP | -5,77% | -3,75% | -5,20% | -4,43% | -12,91% |
| Sharpe | -51,10% | -7,78% | 56,74% | -45,68% | -45,74% |
| Traynor | -43,89% | -9,42% | 47,82% | -50,62% | -17,23% |

Fuente: Elaboración propia (2024)

Dada la información proporcionada en la tabla 14, se destaca que todas las expectativas de retorno de los activos son negativas y tienen volatilidades relativamente altas. Este contexto presenta desafíos en la construcción de un portafolio eficiente. Bajo el supuesto fundamental del modelo, que implica que los inversores buscan maximizar el rendimiento esperado para un nivel específico de riesgo, las expectativas negativas de retorno dificultan la inclusión de dichos activos en la cartera (Perea, A., y Zavaleta, O. H. 2020). En base a lo anterior, se opta por excluir los procesos que demostrarían los resultados matemáticos, ya que en este escenario no se estructura ningún portafolio viable.

**Figura 8.** Frontera eficiente y LMC en el periodo 2020.

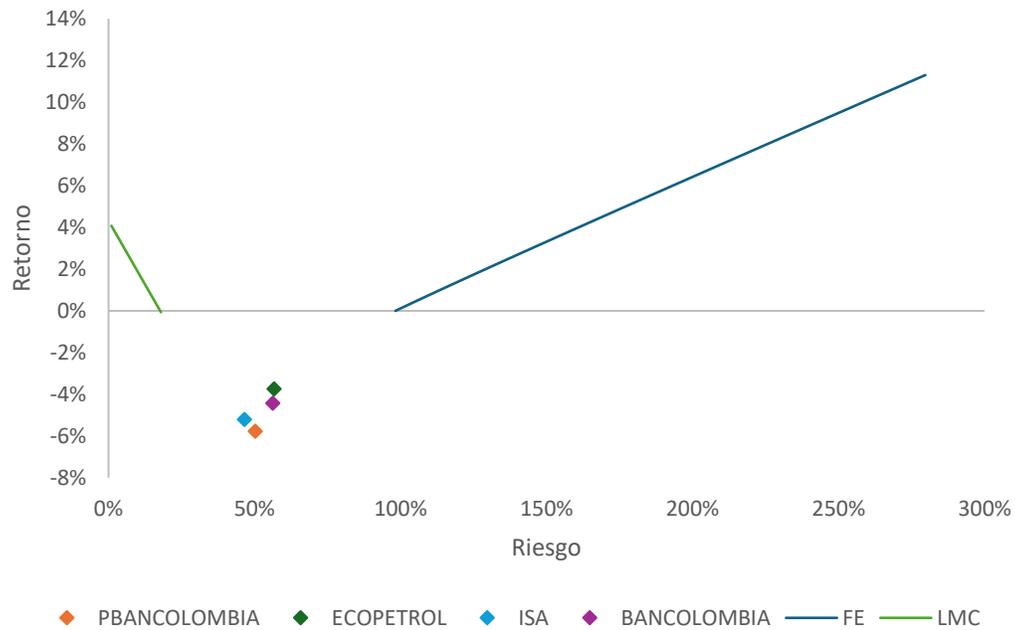

Fuente: Elaboración propia (2024)

**Figura 9.** Comportamiento histórico periodo 2020 - 2023.

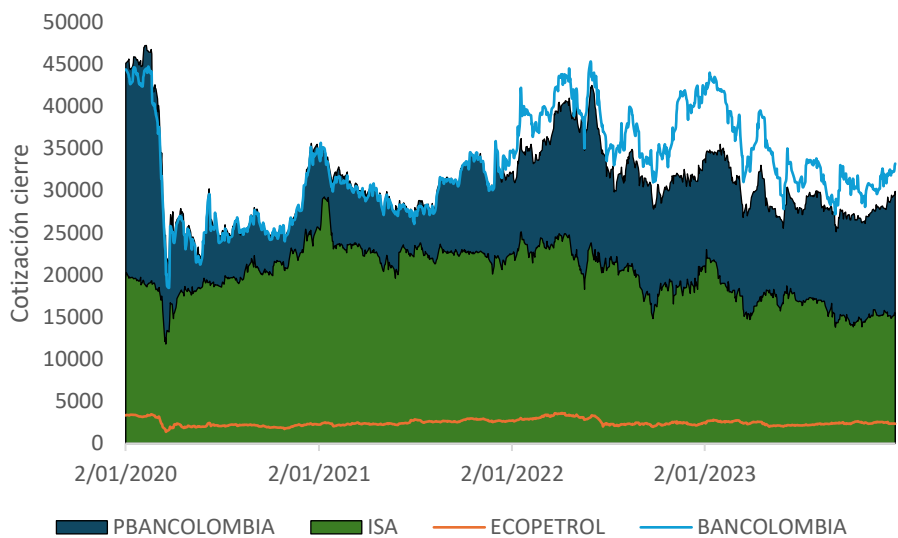

Fuente: Elaboración propia (2024)

**Tabla 15.** Indicadores financieros para el periodo 2020 - 2023.

| Indicador | PBANCOLOMBIA | ECOPETROL | ISA | BANCOLOMBIA | ICOLCAP |
|---|---|---|---|---|---|
| Rendimiento | -7,90% | -6,76% | -4,65% | -5,64% | -6,37% |
| Volatilidad | 35,28% | 43,59% | 42,23% | 42,80% | 26,13% |
| Beta | 49,21% | 34,30% | 26,54% | 35,23% | 100,00% |
| CAMP | 0,56% | 2,59% | 3,65% | 2,46% | -6,37% |
| Sharpe | -42,97% | -32,16% | -28,21% | -30,14% | -52,15% |
| Traynor | -30,81% | -40,88% | -44,89% | -36,62% | -13,63% |

Fuente: Elaboración propia (2024)

**Tabla 16.** Matriz de covarianzas para el periodo 2020 - 2023.

| Covarianza | PBANCOLOMBIA | ECOPETROL | ISA | BANCOLOMBIA |
|---|---|---|---|---|
| PBANCOLOMBIA | 12,44% | 7,70% | 6,64% | 11,09% |
| ECOPETROL | 7,70% | 19,00% | 6,74% | 7,43% |
| ISA | 6,64% | 6,74% | 17,83% | 8,42% |
| BANCOLOMBIA | 11,09% | 7,43% | 8,42% | 18,32% |

Fuente: Elaboración propia (2024)

**Tabla 17.** Matriz inversa de covarianzas para el periodo 2020 - 2023.

| C. Inversa | PBANCOLOMBIA | ECOPETROL | ISA | BANCOLOMBIA |
|---|---|---|---|---|
| PBANCOLOMBIA | 1988% | -371% | -132% | -992% |
| ECOPETROL | -371% | 727% | -132% | -10% |
| ISA | -132% | -132% | 762% | -217% |
| BANCOLOMBIA | -992% | -10% | -217% | 1250% |

Fuente: Elaboración propia (2024)

$$h = \begin{bmatrix} 1 & 1 & 1 & 1 \end{bmatrix} \begin{bmatrix} 1988\% & -371\% & -132\% & -992\% \\ -371\% & 727\% & 132\% & -10\% \\ -132\% & -132\% & 762\% & -217\% \\ -992\% & -10\% & -217\% & 1250\% \end{bmatrix} = \begin{bmatrix} 493\% & 214\% & 281\% & 31\% \end{bmatrix}$$

(1)

$$g = \begin{bmatrix} 0,56\% & 2,5\% & 3,6\% & 2,4\% \end{bmatrix} \begin{bmatrix} 1745\% & -187\% & -22\% & -1240\% \\ -187\% & 502\% & -81\% & -108\% \\ -22\% & -81\% & 762\% & -326\% \\ -1240\% & -108\% & -326\% & 1513\% \end{bmatrix}$$
$$= \begin{bmatrix} -27,7\% & 11,7\% & 18,3\% & 17\% \end{bmatrix}$$

(2)

$$A = \begin{bmatrix} 1 & 1 & 1 & 1 \end{bmatrix} \begin{bmatrix} 493\% \\ 214\% \\ 281\% \\ 31\% \end{bmatrix} = 1020\%$$

(3)

$$B = \begin{bmatrix} 1 & 1 & 1 & 1 \end{bmatrix} \begin{bmatrix} -27,7\% \\ 11,7\% \\ 18,3\% \\ 17\% \end{bmatrix} = 19,3\%$$

$$\gamma = [0{,}56\% \quad 2{,}5\% \quad 3{,}6\% \quad 2{,}4\%] \begin{bmatrix} -27{,}7\% \\ 11{,}7\% \\ 18{,}3\% \\ 17\% \end{bmatrix} = 1{,}24\% \tag{4}$$

$$\delta = 1020\%(1{,}24\%) - 19{,}32\%^2 = 8{,}88\% \tag{5}$$

$$R_t = \frac{(1{,}24\% - 19{,}3\%)7{,}26\%}{(19{,}3\% - 1020\%)7{,}26\%} = 0{,}30\% \tag{6}$$

$$\sigma_{Rt} = \sqrt{\frac{[1020\%(1{,}9\%^2)] - 2(19{,}3\%)(0{,}3\%) + 1{,}24\%}{8{,}8\%}} = 35{,}6\% \tag{7}$$

$$p_{lmc} = \frac{(0{,}3\% - 7{,}26\%)}{35{,}6\%} = -19{,}5\% \tag{8}$$

$$\sigma_{Rt} = \sqrt{\frac{[1020\%(1{,}9\%^2)] - 2(19{,}3\%)(1{,}9\%) + 1{,}24\%}{8{,}8\%}} = 31{,}3\% \tag{9}$$

$$\lambda = \frac{1{,}24\% - 19{,}3\%(1{,}9\%)}{8{,}88\%} = 9{,}7\% \tag{10}$$

$$\theta = \frac{1020\%(1{,}9\% - 19{,}3\%)}{8{,}88\%} = 0{,}64\% \tag{11}$$

$$\omega = 9{,}7\%[493\% \quad 214\% \quad 281\% \quad 31\%] + 0{,}64\%[-27{,}7\% \quad 11{,}7\% \quad 18{,}3\% \quad 17\%] \tag{12}$$

**Tabla 18.** Cálculo de los pesos a partir de la ecuación 13.

| Nemotécnico | Formula | Peso |
|---|---|---|
| PBANCOLOMBIA | 9,7%(493%) + 0,64%(−27,7%) | 48% |
| ECOPETROL | 9,7%(214%) + 0,64%(11,7%) | 21% |
| ISA | 9,7%(281%) + 0,64%(18,3%) | 28% |
| BANCOLOMBIA | 9,7%(31%) + 0,64%(17%) | 3% |
| | Total | 100% |

Fuente: Elaboración propia (2024)

El retorno del portafolio se puede reconfirmar con la suma producto de $\omega$ y $E(R)$ lo siguiente:

$$n = \begin{bmatrix} 0{,}56\% \\ 2{,}5\% \\ 3{,}6\% \\ 2{,}4\% \end{bmatrix} \begin{bmatrix} 48\% \\ 21\% \\ 28\% \\ 3\% \end{bmatrix} = \begin{bmatrix} 0{,}27\% \\ 0{,}55\% \\ 1{,}01\% \\ 0{,}08\% \end{bmatrix} \tag{13}$$

$$\tag{14}$$

$$R_{p2} = \sum_{i=1}^{n=4} n = 1{,}9\%$$

(15)

$$v = \begin{bmatrix} 48\% & 21\% & 28\% & 3\% \end{bmatrix} \begin{bmatrix} 12{,}44\% & 7{,}7\% & 6{,}64\% & 11{,}09\% \\ 7{,}70\% & 19\% & 6{,}74\% & 7{,}43\% \\ 6{,}64\% & 6{,}74\% & 17{,}83\% & 8{,}42\% \\ 11{,}09\% & 7{,}43\% & 8{,}42\% & 18{,}32\% \end{bmatrix} \begin{bmatrix} 48\% \\ 21\% \\ 28\% \\ 3\% \end{bmatrix} = 9{,}8\%$$

(16)

$$\sigma_i^2 = \sqrt{9{,}8\%} = 31{,}3\%$$

(17)

$$Sh = \frac{1{,}9\% - 7{,}2\%}{31{,}3\%} = -17{,}1\%$$

(18)

**Figura 10.** Frontera eficiente y LMC en el periodo 2020 – 2023.

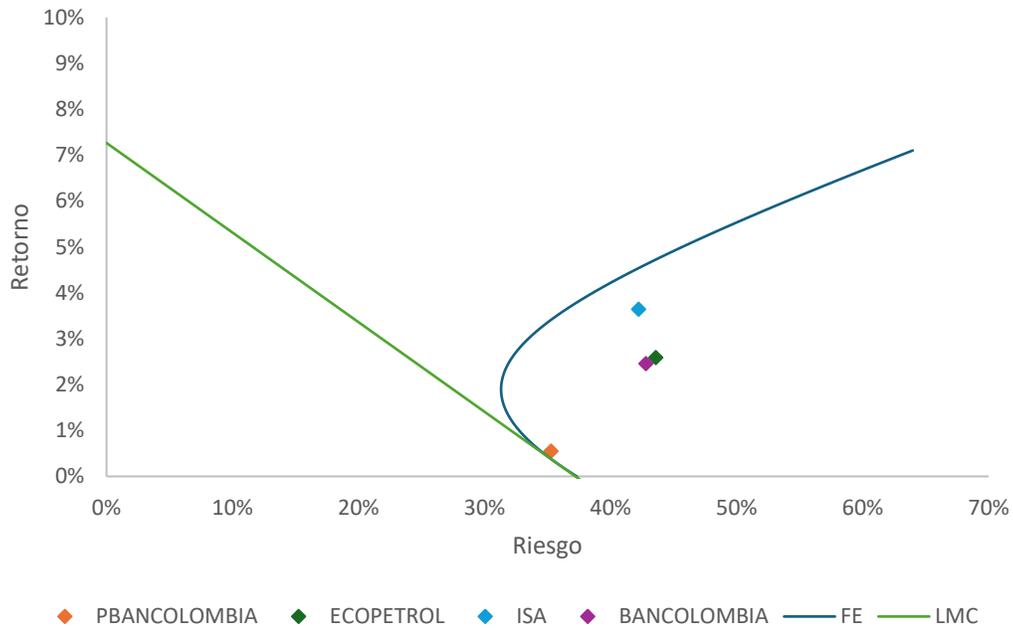

Fuente: Elaboración propia (2024)

**Figura 11.** Comportamiento periodo 2023.

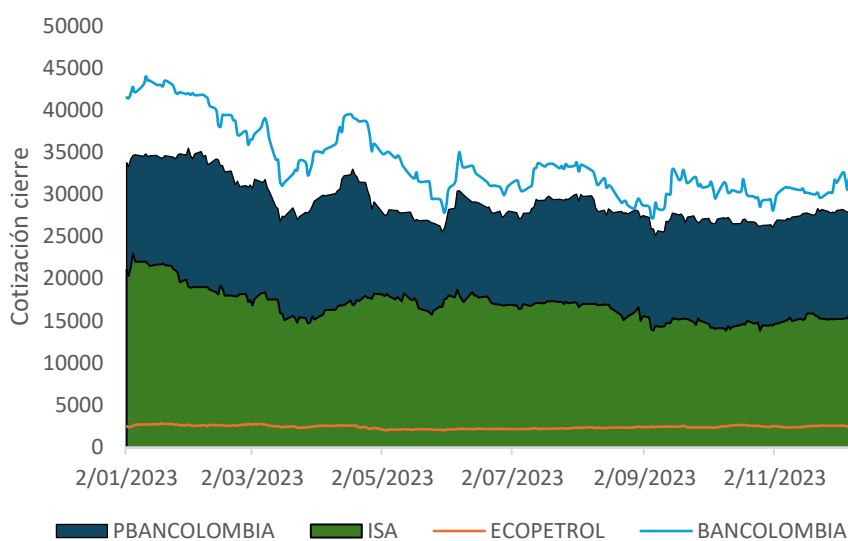

Fuente: Elaboración propia (2024)

**Tabla 19.** Indicadores financieros.

| Indicador | PBANCOLOMBIA | ECOPETROL | ISA | BANCOLOMBIA | ICOLCAP |
|---|---|---|---|---|---|
| Rendimiento | -11,35% | -5,61% | -26,57% | -20,00% | -6,92% |
| Volatilidad | 27,51% | 33,55% | 40,06% | 38,41% | 18,29% |
| Beta | 29,86% | 23,22% | 11,75% | 17,10% | 100,00% |
| CAMP | 4,91% | 6,03% | 7,96% | 7,06% | -6,92% |
| Sharpe | -77,41% | -46,36% | -91,14% | -77,96% | -92,23% |
| Traynor | -71,32% | -66,98% | -310,64% | -175,14% | -16,87% |

Fuente: Elaboración propia (2024)

**Tabla 20.** Matriz de covarianzas.

| Covarianza | PBANCOLOMBIA | ECOPETROL | ISA | BANCOLOMBIA |
|---|---|---|---|---|
| PBANCOLOMBIA | 7,57% | 4,16% | 3,40% | 4,93% |
| ECOPETROL | 4,16% | 11,26% | 2,72% | 2,34% |
| ISA | 3,40% | 2,72% | 16,05% | 5,46% |
| BANCOLOMBIA | 4,93% | 2,34% | 5,46% | 14,76% |

Fuente: Elaboración propia (2024)

**Tabla 21.** Matriz inversa de covarianzas.

| C. Inversa | PBANCOLOMBIA | ECOPETROL | ISA | BANCOLOMBIA |
|---|---|---|---|---|
| PBANCOLOMBIA | 2084% | -621% | -151% | -542% |
| ECOPETROL | -621% | 1125% | -79% | 58% |
| ISA | -151% | -79% | 740% | -211% |
| BANCOLOMBIA | -542% | 58% | -211% | 928% |

Fuente: Elaboración propia (2024)

$$h = \begin{bmatrix} 1 & 1 & 1 & 1 \end{bmatrix} \begin{bmatrix} 2084\% & -621\% & -151\% & -542\% \\ -621\% & 1125\% & -79\% & 58\% \\ -151\% & -79\% & 740\% & -211\% \\ -542\% & 58\% & -211\% & 928\% \end{bmatrix} = \begin{bmatrix} 770\% & 483\% & 299\% & 233\% \end{bmatrix}$$

(1)

$$g = \begin{bmatrix} 4,9\% & 6\% & 7,9\% & 7\% \end{bmatrix} \begin{bmatrix} 1745\% & -187\% & -22\% & -1240\% \\ -187\% & 502\% & -81\% & -108\% \\ -22\% & -81\% & 762\% & -326\% \\ -1240\% & -108\% & -326\% & 1513\% \end{bmatrix}$$
$$= \begin{bmatrix} 14,5\% & 35,1\% & 31,8\% & 25,6\% \end{bmatrix}$$

(2)

$$A = \begin{bmatrix} 1 & 1 & 1 & 1 \end{bmatrix} \begin{bmatrix} 770\% \\ 483\% \\ 299\% \\ 233\% \end{bmatrix} = 1785\%$$

(3)

$$B = \begin{bmatrix} 1 & 1 & 1 & 1 \end{bmatrix} \begin{bmatrix} 14,5\% \\ 35,1\% \\ 31,8\% \\ 25,6\% \end{bmatrix} = 107\%$$

(4)

$$\gamma = \begin{bmatrix} 4,9\% & 6\% & 7,9\% & 7\% \end{bmatrix} \begin{bmatrix} 14,5\% \\ 35,1\% \\ 31,8\% \\ 25,6\% \end{bmatrix} = 7,2\%$$

(5)

$$\delta = 1785\%(7,2\%) - 107\%^2 = 13,3\%$$

(6)

$$R_t = \frac{(7,2\% - 107\%)9,95\%}{(107\% - 1785\%)9,95\%} = 4,9\%$$

(7)

$$\sigma_{Rt} = \sqrt{\frac{[1020\%(4,9\%^2) - 2(107\%)(4,9\%) + 7,2\%]}{4,9\%}} = 26,7\%$$

(8)

$$p_{lmc} = \frac{(4,9\% - 9,95\%)}{26,7\%} = -18,8\%$$

(9)

$$\sigma_{Rt} = \sqrt{\frac{[1785\%(6\%^2) - 2(107\%)(6\%) + 7,2\%]}{13,3\%}} = 23,6\%$$

(10)

$$\lambda = \frac{7,2\% - 107\%(6\%)}{13,3\%} = 5,6\%$$

(11)

$$\theta = \frac{1785\%(6\% - 107\%)}{13,3\%} = -0,75\%$$

(12)

$$\omega = 5,6\%\begin{bmatrix} 770\% & 483\% & 299\% & 233\% \end{bmatrix} + -0,75\%\begin{bmatrix} 14,5\% & 35,1\% & 31,8\% & 25,6\% \end{bmatrix}$$

(13)

**Tabla 22.** Cálculo de los pesos a partir de la ecuación 13.

| Nemotécnico | Formula | Peso |
|---|---|---|
| PBANCOLOMBIA | 5,6%(770%) + −0,75%(14,5%) | 43% |
| ECOPETROL | 5,6%(483%) + −0,75%(35,1%) | 27% |
| ISA | 5,6%(299%) + −0,75%(31,8%) | 17% |
| BANCOLOMBIA | 5,6%(233%) + −0,75%(25,6%) | 13% |
| Total | | 6,0% |

Fuente: Elaboración propia (2024)

El retorno del portafolio se puede reconfirmar con la suma producto de $\omega$ y $E(R)$ lo siguiente:

$$n = \begin{bmatrix} 43\% \\ 27\% \\ 17\% \\ 13\% \end{bmatrix} \begin{bmatrix} 4,9\% \\ 6,0\% \\ 7,9\% \\ 7,0\% \end{bmatrix} = \begin{bmatrix} 2,13\% \\ 1,63\% \\ 1,33\% \\ 0,92\% \end{bmatrix}$$

(14)

$$R_{p2} = \sum_{i=1}^{n=4} n = 6,0\%$$

(15)

$$v = \begin{bmatrix} 43\% & 27\% & 17\% & 13\% \end{bmatrix} \begin{bmatrix} 7,57\% & 4,16\% & 3,40\% & 4,93\% \\ 4,16\% & 11,2\% & 2,72\% & 2,34\% \\ 3,40\% & 2,72\% & 16\% & 5,46\% \\ 4,93\% & 2,34\% & 5,46\% & 14,7\% \end{bmatrix} \begin{bmatrix} 43\% \\ 27\% \\ 17\% \\ 13\% \end{bmatrix} = 5,6\%$$

(16)

$$\sigma_i^2 = \sqrt{5,6\%} = 23,6\%$$

(17)

$$Sh = \frac{6,0\% - 9,9\%}{23,6\%} = -16,6\%$$

(18)

**Figura 12.** Frontera eficiente y LMC en el periodo 2023.

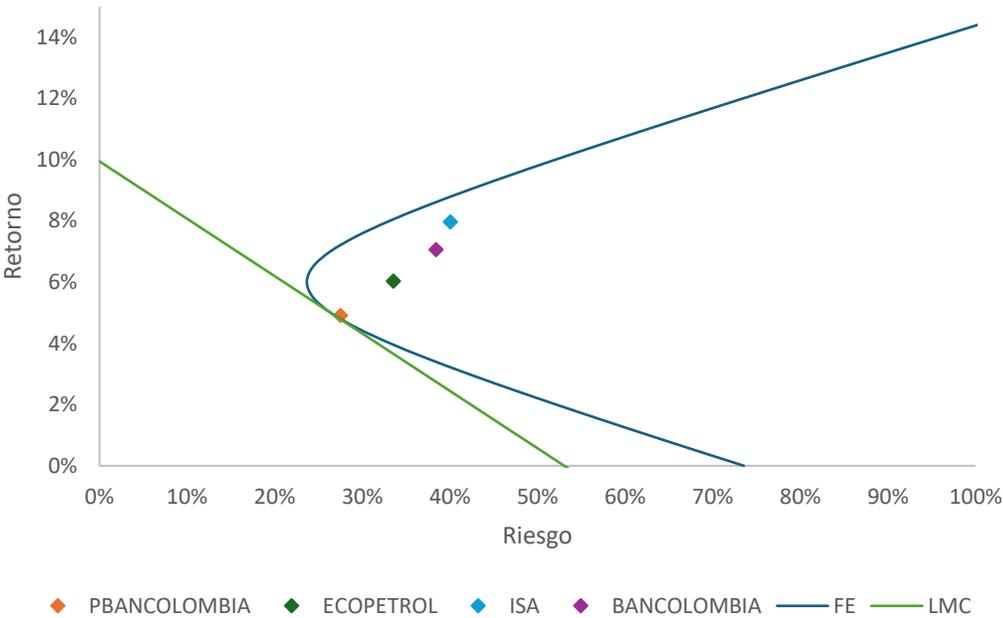

Fuente: Elaboración propia (2024).

**Síntesis Resultados**

**Tabla 23.** Resumen de comportamiento de los portafolios por periodos.

| Indicador | 2015 - 2023 | 2015 - 2019 | 2016 – 2020 | 2020 - 2023 | 2023 |
|---|---|---|---|---|---|
| Rendimiento | 3,80% | 5,70% | 5,40% | 1,90% | 6,00% |
| Beta | 0,83 | 0,61 | 1,02 | 0,91 | 0,49 |
| Varianza | 6,41% | 3,41% | 6,07% | 9,80% | 5,60% |
| Riesgo | 25,31% | 18,45% | 24,64% | 31,31% | 23,67% |
| Sharpe | -12,13% | -4,60% | -2,19% | -17,13% | -16,68% |

Fuente: Elaboración propia (2024).

**Tabla 24.** Resumen de pesos por periodos.

| Nemotécnico | 2015 - 2023 | 2015 - 2019 | 2016 – 2020 | 2020 - 2023 | 2023 |
|---|---|---|---|---|---|
| PBANCOLOMBIA | 49,81% | 30,89% | 16,03% | 48,10% | 43,36% |
| ECOPETROL | 17,12% | 15,47% | 25,01% | 21,06% | 27,03% |
| ISA | 31,31% | 43,37% | 52,50% | 27,67% | 16,65% |
| BANCOLOMBIA | 1,76% | 10,28% | 6,45% | 3,16% | 12,96% |

Fuente: Elaboración propia (2024).

**Tabla 25.** Resumen retornos por periodo.

|  | Nemotécnico | 2015 - 2023 | 2015 - 2019 | 2016 – 2020 | 2020 - 2023 | 2023 |
|---|---|---|---|---|---|---|
| Histórico | PBANCOLOMBIA | 0,81% | 10,48% | 10,20% | -7,90% | -11,35% |
| | ECOPETROL | 2,86% | 12,80% | 15,13% | -6,76% | -5,61% |
| | ISA | 7,72% | 19,84% | 28,34% | -4,65% | -26,57% |
| | BANCOLOMBIA | 2,21% | 10,05% | 10,77% | -5,64% | -20,00% |
| CAPM | PBANCOLOMBIA | 3,03% | 5,43% | 5,27% | 0,56% | 4,91% |
| | ECOPETROL | 4,32% | 5,82% | 5,48% | 2,59% | 6,03% |
| | ISA | 4,73% | 5,89% | 5,41% | 3,65% | 7,96% |
| | BANCOLOMBIA | 3,92% | 5,54% | 5,35% | 2,46% | 7,06% |
| Markowitz | PBANCOLOMBIA | 1,51% | 1,68% | 0,84% | 0,27% | 2,13% |
| | ECOPETROL | 0,74% | 0,90% | 1,37% | 0,55% | 1,63% |
| | ISA | 1,48% | 2,56% | 2,84% | 1,01% | 1,33% |
| | BANCOLOMBIA | 0,07% | 0,57% | 0,35% | 0,08% | 0,92% |

Fuente: Elaboración propia (2024).

# 4. CONCLUSIONES

Las carteras de varianza mínima, orientadas a invertir en acciones con baja volatilidad, han demostrado un sólido desempeño a lo largo de la historia. Este rendimiento positivo se correlaciona con la observación empírica de que las acciones con baja beta, es decir, menos sensibles al mercado, tienden a exhibir rendimientos más elevados.

Esta correlación se ilustra en la tabla 23, donde durante el periodo 2015-2019, una beta más baja (0,61) se asocia con un rendimiento más alto (5,70%), mientras que en el periodo 2016-2020, una beta más alta (1,02) está vinculada a un rendimiento ligeramente menor (5,40%). Según Ang et al. (2006; 2009), esta relación se explica parcialmente por la "anomalía de riesgo idiosincrásico", indicando que acciones con mayor riesgo específico de la empresa tienden a generar rendimientos más elevados.

En los periodos mencionados, se observa una reducción del rendimiento de 0,30 puntos porcentuales (p.p), acompañada por un aumento en la volatilidad de 6,19 p.p, respaldado por el incremento en los valores de beta. En el periodo 2020-2023, marcado por la fase pandémica, estos resultados se agudizan, con un retorno del portafolio del 1,90% y una volatilidad del 31,3%, niveles de riesgo no alcanzados anteriormente.

En el periodo de recuperación en 2023, se registra un retorno del 6%, con una volatilidad del 23,6%. Sin embargo, es importante mencionar que ninguno de estos portafolios logra compensar adecuadamente el riesgo asumido.

De manera notable, se destaca que PBANCOLOMBIA tiene la mayor ponderación en el histórico de los portafolios (tabla 24), ya que en su mayoría se ubica en la tangente de la línea de frontera eficiente (FE), considerándose un activo "óptimo". De acuerdo con Markowitz (1952), esto implica que ofrece la mejor relación entre riesgo y retorno posible, es decir, la máxima rentabilidad por unidad de riesgo.